\documentclass[apj]{emulateapj}

\usepackage{amsmath}
\usepackage{graphicx}
\usepackage{graphics}
\usepackage{epsf,psfig}
\usepackage[dvips]{color}
\usepackage{color}
\usepackage{epsfig}
\usepackage{amssymb}
\usepackage{amsfonts}
\usepackage{latexsym}
\usepackage{txfonts}

\newcommand\xte{{\it RXTE\/}}

\newcommand\chandra{{\it Chandra}}

\newcommand\integ{{\it INTEGRAL\/}}

\newcommand\ibisg{IBIS/ISGRI}
\newcommand\swi{{\it Swift\/}}

\newcommand\fermi{{\it Fermi\/}}

\newcommand\hess{{H.E.S.S.}}

\newcommand\igr{IGR~J14003$-$6326}
\newcommand\psr{PSR~J1400$-$6325}
\newcommand\snr{G310.6$-$1.6}

\newcommand{\un}[1]{~\hspace{-1pt}\ensuremath{\mathrm{#1}}}
\newcommand{\ti}[1]{$^{44}{}${#1}}

\newcommand{\gammarays}{$\gamma$-rays}
\newcommand{\xray}{X-ray}
\newcommand{\xrays}{X-rays}

\newcommand{\be}{\begin{equation}}
\newcommand{\ee}{\end{equation}}
\newcommand{\ben}{\begin{eqnarray}}
\newcommand{\een}{\end{eqnarray}}
\newcommand{\bc}{\begin{center}}
\newcommand{\ec}{\end{center}}

\def\amin{\ifmmode^{\prime}\else$^{\prime}$\fi}
\def\asec{\ifmmode^{\prime\prime}\else$^{\prime\prime}$\fi}
\def\d{$^\circ$}
\def\m{$^{\prime}$}
\def\s{$^{\prime\prime}$}
\def\hh{$^{\mathrm h}$}
\def\mm{$^{\mathrm m}$}
\def\sec{$^{\mathrm s}$}

\def\edot{$\dot{{\rm E}}$}

\def\kms{km\,s$^{-1}$}
\def\cm3{cm$^{-3}$}

\def\msun{$M_\odot$}
\def\nh{N$_{\rm H}$}

\def\eg{{\it e.g.~}}
\def\etal{et~al.~}
\def\ie{{\em i.e.~}}

\slugcomment{Accepted for publication in ApJ}

\shorttitle{Discovery of \psr~associated with \igr}
\shortauthors{Renaud et al.}

\begin{document}

\title{Discovery of a highly energetic pulsar associated with \igr~in \\ 
       a young uncataloged Galactic supernova remnant \snr}

\author{M.~Renaud\altaffilmark{1,2}, V.~Marandon\altaffilmark{1}, E.~V.~Gotthelf\altaffilmark{3}, 
	J.~Rodriguez\altaffilmark{4}, R.~Terrier\altaffilmark{1}, F.~Mattana\altaffilmark{1}, 
	F.~Lebrun\altaffilmark{1}, J.~A.~Tomsick\altaffilmark{5} and R.~N.~Manchester\altaffilmark{6}}
\email{mrenaud@lpta.in2p3.fr}
\altaffiltext{1}{AstroParticule et Cosmologie (APC), CNRS-UMR 7164, Universit\'e Paris 7 Denis Diderot, F-75205 Paris, France}
\altaffiltext{2}{Laboratoire de Physique Th\'eorique et Astroparticules (LPTA), CNRS-UMR 5207, Universit\'e Montpellier II, F-34095 Montpellier, France}
\altaffiltext{3}{Columbia Astrophysics Laboratory, Columbia University, 550 West 120$^{th}$ Street, New York, NY 10027, USA}
\altaffiltext{4}{CEA Saclay, Laboratoire AIM, CNRS-UMR 7158, DSM/IRFU/Service d'Astrophysique, F-91191 Gif-sur- Yvette, France}
\altaffiltext{5}{Space Sciences Laboratory, 7 Gauss Way, University of California, Berkeley, CA 94720-7450, USA}
\altaffiltext{6}{CSIRO Astronomy and Space Science, Australia Telescope National Facility, P.O. Box 76, Epping NSW 1710, Australia}

\begin{abstract}

We report the discovery of 31.18\un{ms} pulsations from the \integ~source \igr~using the Rossi X-ray 
Timing Explorer (\xte). This pulsar is most likely associated with the bright Chandra \xray~point source 
lying at the center of \snr, a previously unrecognised Galactic composite supernova remnant with a bright central 
non-thermal radio and \xray~nebula, taken to be the pulsar wind nebula (PWN). \psr~is amongst the most 
energetic rotation-powered pulsars in the Galaxy, with a spin-down luminosity of $\dot E = 5.1 \times 10^{37}$ 
erg~s$^{-1}$. In the rotating dipole model, the surface dipole magnetic field strength is 
$B_s = 1.1 \times10^{12}$~G and the characteristic age $\tau_c \equiv P/2\dot P = 12.7$~kyr. The high spin-down
power is consistent with the hard spectral indices of the pulsar and the nebula of $1.22\pm0.15$ and $1.83\pm0.08$, 
respectively, and a 2--10\un{keV} flux ratio $F_{PWN}/F_{PSR} \sim 8$. Follow-up Parkes observations resulted 
in the detection of radio emission at 10 and 20\un{cm} from \psr~at a dispersion measure of $\sim$ 560 cm$^{-3}$ pc, which 
implies a relatively large distance of 10 $\pm$ 3\un{kpc}. However, the resulting location off the Galactic Plane  
of $\sim$ 280\un{pc} would be much larger than the typical thickness of the molecular disk, and we argue that \snr~lies
at a distance of $\sim$ 7\un{kpc}. There is no gamma-ray counterpart to the nebula or pulsar in the Fermi data published 
so far. A multi-wavelength study of this new composite supernova remnant, from radio to very-high energy gamma-rays, 
suggests a young ($\lesssim$ 10$^{3}$ yr) system, formed by a sub-energetic ($\lesssim 10^{50}$ ergs), low ejecta mass 
(M$_{\rm ej} \sim 3$ \msun) SN explosion that occurred in a low-density environment ($n_0 \sim$ 0.01 cm$^{-3}$). 

\end{abstract}


\section{Introduction}
\label{s:intro}

The number of known supernova remnants (SNRs) in the Galaxy has increased significantly over the 
last few years, mainly due to a new generation of radio and \xray~instruments of unprecedented
sensitivities and angular resolutions. In particular, these observations, combined with Galactic 
Plane surveys \cite[\eg][]{c:brogan06}, and targeted observations \cite[\eg][]{c:gelfand07,c:gaensler08} 
have increased the fraction of SNRs found to harbor an energetic pulsar powering a wind nebula (the 
so-called composite SNRs). Moreover, deep observations toward compact PWNe have proven to be successful 
in detecting their powering pulsars \citep{c:camilo09,c:gotthelf09}, providing constraints on their 
energetics, spin evolution, and birth parameters. Along with the current generation of high/very-high 
energy (VHE; $>$ 100\un{GeV}) instruments, these observations are of prime importance for understanding 
the structure and evolution of these sources, and the underlying acceleration mechanisms which occur 
close to the pulsar, and at the relativistic and non-relativistic shock fronts bounding the PWN and
the host SNR \citep{c:gs06}. 

The soft $\gamma$-ray source \igr~was discovered in a deep \integ/\ibisg~mosaic of the Circinus region as 
a persistent source at the mCrab level \citep{c:keek06}. A \swi/XRT survey of \integ~sources located 
\igr~to 4\s~in 2--10\un{keV} \xrays, but no conclusion was reached on its nature \citep{c:malizia07}. 
A follow-up \chandra~survey of unidentified IGR sources reported a PWN within a $\sim$ 3\m~diameter 
nearly circular emission nebula \citep{c:tomsick09}. These authors presented the 0.3--10\un{keV} spectrum of 
the total emission by an absorbed power-law with a relatively hard photon index ($\Gamma$ = 1.82 $\pm$ 0.13) 
and a large column density \nh~$\sim$ 3 $\times$ 10$^{22}$ cm$^{-2}$. 

In this paper, we report the discovery with the Rossi X-ray Timing Explorer (\xte) of 31.18\un{ms} pulsations 
toward \igr, and the radio detection in follow-up observations using the Parkes telescope. We present a 
multi-wavelength study of this new Galactic composite SNR, \snr, using radio, X-ray, and gamma-ray data. \psr~is 
one of the most energetic in the Galaxy and powers a wind nebula whose broadband non-thermal synchrotron emission 
is measured in radio and \xrays. We present a spatial and spectral analysis that shows evidence for a 
young SNR, formed from a sub-energetic, low ejecta mass SN explosion that occurred in a low-density environment. 
We also discuss the implications of the lack of high-energy (HE) gamma-ray emission in the \fermi~data.

\section{Observations and Results}
\label{s:obs}

\subsection{\chandra}
\label{s:chandra}

\begin{figure*}[!htb]
\begin{center}

\includegraphics[width=0.80\linewidth,angle=0]{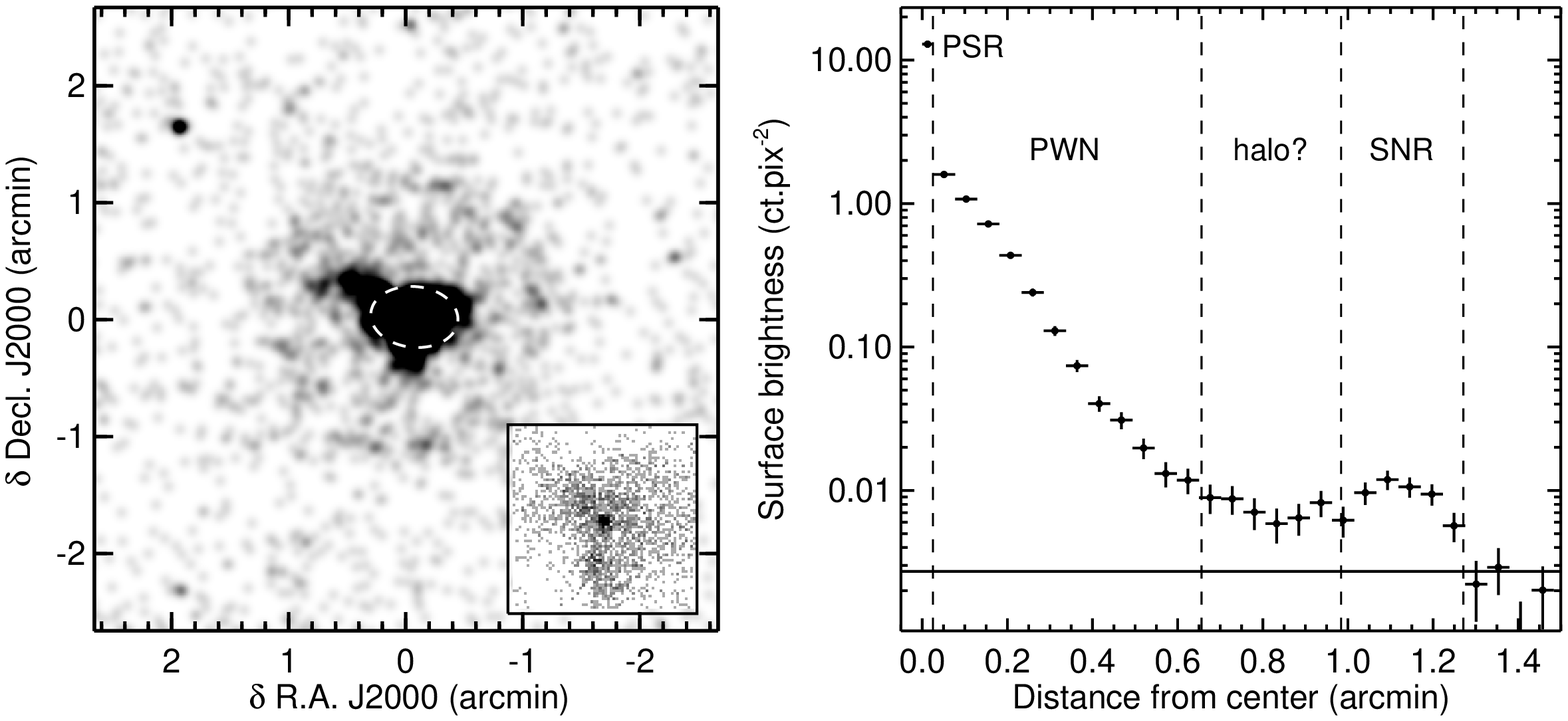}

\caption{{\it Left:} \chandra/ACIS-I count rate image of \snr~in the 0.5--10\un{keV} energy range, smoothed 
with a gaussian of $\sigma$ = 5\s~(linear scale). The dashed ellipse shows the intrinsic size of the 
radio MGPS-2 extended source cataloged at 843\un{MHz} \citep[][see section \ref{s:radio}]{c:murphy07}. 
The bright, central, point-like source, is better revealed in the inset unsmoothed image of the central 
regions of the SNR. {\it Right:} Surface brightness profile of the \chandra/ACIS emission, centered on the 
point-like source. The background level is depicted by the horizontal solid line. The dashed lines delineate 
the regions of interest, namely the PSR, the PWN, the putative \xray~halo, and the host SNR. \label{f:ima_chandra}}

\end{center}
\end{figure*}

We have performed a reanalysis of the \chandra~data of the \igr~field \citep{c:tomsick09} taken on 
2008 June 29 UT, using the Advanced CCD Imaging Spectrometer \citep[ACIS,][]{c:garmire03} and operating in 
TIME/FAINT exposure mode. The livetime amounts to 5077\un{s}, corrected from the 1.3\% readout dead time. 
The maximum pixel count rate within the source is 0.015 ct s$^{-1}$, so pile-up effect could be safely ignored. 
All the data reduction and analysis were performed with the \chandra~Interactive Analysis of Observation software 
(CIAO) version 4.1.2, using the CALibration DataBase (CALDB) v4.1.3. Figure \ref{f:ima_chandra} (left) 
shows the 0.5--10\un{keV} \chandra~image centered on \igr, which exhibits a characteristic morphology of 
a composite SNR, as revealed by the radial profile of the surface brightness (Figure \ref{f:ima_chandra}, 
right). In particular, the excess emission at 1.1-1.2\m~from the center is found to be the clear 
signature of the host shell-type SNR. We have considered three regions of interest, namely the bright 
point-like source (PSR), the diffuse surrounding emission (PWN), and the shell-like structure (SNR). As shown 
on the profile, some faint emission seems to extend up to the shell-type structure, at distances of $\sim$ 
1\m~from the PSR. In view of the large \nh, this could be partly explained by a dust-scattering halo, similar to 
those observed in other PWNe, such as G21.5-0.9 \citep{c:slane00,c:safi-harb01,c:bocchino05}. Unfortunately, 
the lack of statistics prevents us from drawing any firm conclusion about the nature of this extended 
\xray~emission. The source-free region used to extract the background was taken on the same CCD chip as the 
source, in a circle of 1.45\m~in radius. Apart from the PWN emission, the low count levels led us to fit 
the three components simultaneously, with a common hydrogen column density \nh~= (2.09 $\pm$ 0.12) $\times$ 
10$^{22}$ cm$^{-2}$. As noticed by \citet{c:tomsick09}, this value is of the same order as the Galactic 
value\footnote{We here make use of the \citet{c:anders89} abundances, rather those of \citet{c:wilms00}, 
as done in \citet{c:tomsick09}. Using the latter abundances yields compatible \nh~$\sim$ 3 $\times$ 
10$^{22}$ cm$^{-2}$.}, and suggests a large distance and/or local absorption. 

The best-fit position of the point-like source is R.A.(J2000) = 14\hh00\mm45.69\sec, Decl.(J2000) = 
$-$63\d25\m42.6\s~\citep{c:tomsick09}. For the spectral analysis, we chose an integration region of 1.5\s~in
radius centered on the source and estimated the source background using counts extracted from a concentric 
annulus of radii 1.5\s~and 2\s. Fits to the background subtracted spectrum are not tightly constrained, even 
though a power-law gives slightly better results than any thermal model. The spectrum is hard, with a best-fit 
power-law index of 1.22 $\pm$ 0.15 and an unabsorbed 2--10\un{keV} flux of (1.95 $\pm$ 0.5) $\times$ 10$^{-12}$ 
ergs cm$^{-2}$ s$^{-1}$. The PWN integration region was taken in an annulus centered on the PSR position, 
with an inner radius of 1.5\s~and an outer radius of 0.65\m~(see Fig. \ref{f:ima_chandra}, right). The 
best-fit power-law index and the unabsorbed 2--10\un{keV} flux are found to be 1.83 $\pm$ 0.09 and (1.51 $\pm$
0.20) $\times$ 10$^{-11}$ ergs cm$^{-2}$ s$^{-1}$, respectively. Therefore, the PWN is by far the dominant
component in \xrays~and these spectral results are in close agreement with those obtained by \citet{c:tomsick09}
on the total emission. The SNR integration region was defined as an annulus, with inner and outer radii of 0.98 
and 1.27\m, respectively. As for the PSR, due to the low statistics, the SNR spectrum is rather poorly constrained, 
and does not allow us to clearly distinguish between thermal and power-law models, although no strong \xray~lines 
are present. A power-law fit gives a rather soft spectrum, with a best-fit spectral index of 2.56 $\pm$ 0.18 and an 
unabsorbed 2--10\un{keV} flux of (1.0 $\pm$ 0.2) $\times$ 10$^{-12}$ ergs cm$^{-2}$ s$^{-1}$. Whatever is the nature 
of the \xray~emission, the flux measured with \chandra~can be considered as an upper limit on the synchrotron emission. 
We then converted the best-fit power-law spectrum to a surface brightness at 4\un{keV} of $\sim$ 5 $\times$ 10$^{32}$ 
ergs pc$^{-2}$ s$^{-1}$ sr$^{-1}$, as this will be compared to theoretical predictions in section \ref{s:discu}.

\subsection{\xte}
\label{s:rxte}

The \xray~flux measured from the \chandra~point source in \snr~is sufficient to search for the expected 
pulsations using \xte. A pair of 44~ks exposures were obtained a year apart on 2008 September 29 UT and 
2009 September 30. Data were collected with the Proportional Counter Array \citep[PCA;][]{c:jah96} in the 
GoodXenon mode with an average of 1.5 and 2.2 out of the five proportional counter units (PCUs) active, 
for the two observations, repectively. In this mode, photons are time-tagged to $0.9$~$\mu$s and have an 
absolute uncertainty better than 100~$\mu$s. The effective area of five combined units is $6500$ cm$^{2}$ 
at 10\un{keV} with a roughly circular field-of-view of $\sim$ 1\d~FWHM. Spectral information 
is available in the 2--60\un{keV} energy band with a resolution of $\sim$ 16\% at 6\un{keV}. The standard 
time filters were applied to the PCA data, which rejects intervals of South Atlantic Anomaly passages, 
Earth occultations, and other periods of high particle activity. The photon arrival times were transformed 
to the Solar-system barycenter in Barycentric Dynamical Time (TDB) using the JPL DE200 ephemeris and the 
\chandra~coordinates for the point source.

\subsubsection{Timing analysis}

\begin{figure}[!htb]
\hspace{0.2cm}
\begin{center}

\includegraphics[height=0.88\linewidth,angle=270,clip=true]{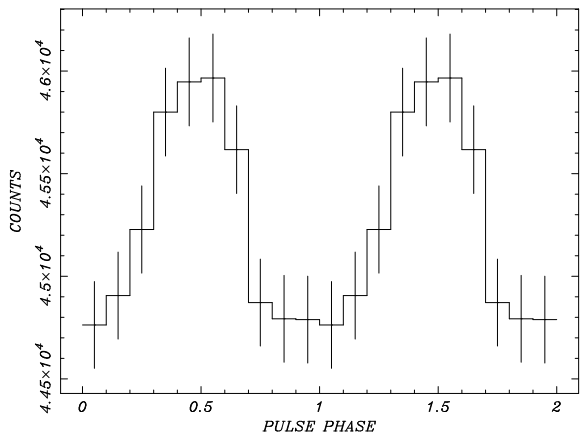}

\caption{\psr~folded light curve in the \xte~PCA 2--20\un{keV} band. Two cycles 
         are shown for clarity. Phase zero is arbitrary. \label{f:rxte1}}

\end{center}
\end{figure}

We restricted the \xte~timing analysis to photon data with energies in the 2--20\un{keV} range (PCA channels 
2--50) recorded in the top Xenon layer of each PCU, to optimize the signal-to-noise ratio. For the first 
observation, we further excluded one 4\un{ks} non-contiguous data segment at the beginning of the run, 
resulting in 33.4\un{ks} of filtered time spanning 45\un{ks}. A $2^{26}$ Fast Fourier Transform (FFT) of the 
data binned in 1\un{ms} steps revealed a significant signal of power $S = 55$ at a period of $P = 31.1$\un{ms}. 
For the $6.7 \times 10^7$ search elements this corresponds to a false detection probability for a blind search 
down to 1\un{ms} of $\wp = 7.6 \times 10^{-5}$ (99.9924 \% Confidence Level). We refined this signal using 
a $Z^2_1$ statistics which yields a signal power $Z^2_1 = 53.76$ at a period of $P=31.180373(2)$\un{ms}. The 
$1\sigma$ uncertainty on the last digit is given in parentheses. Figure~\ref{f:rxte1} displays the pulse profile 
in the 2--20\un{keV} band folded at this period; the profile is narrower than a sinusoid, somewhat boxy, 
with a well deliniated off-pulse region. We do not see any evidence for energy dependence of the pulse profile 
when subdividing the 2--20\un{keV} band. The signal does not match any cataloged pulsar in the field-of-view, 
in particular the pulsars PSR~J1403$-$6310 ($P=0.39910$\un{s}) and PSR~J1401$-$6357 ($P=0.84279$\un{s}). In order 
to ensure that we have not picked up an aliased signal from these PSRs, the data was rebarycentered at their 
respective coordinates and searched for the known pulse period. No signal was detected apart from the original 
one reported above.

Based on a probable detection of an highly energetic PSR, we requested a second \xte~observation to confirm 
this result and determine the PSR spin-down rate. For the 2009 data, we followed the same proceedures outlined 
above and constructed a $Z^2_1$ periodogram around the expected signal. This yielded a significant signal at 
$P=31.1816011(3)$\un{ms} with power $Z^2_1 = 116.28$, corresponding to a negligible probability of false 
detection. This result confirms the previous detection and provides a period derivative measurement of 
$\dot P = (3.890\pm0.006) \times 10^{-14}$. The one sigma error is derived from the root sum of the variance of 
the period uncertainties. The ephemeris is presented in Table~\ref{t:ephem} along with the inferred spin-down 
parameters. The derived spin-down power \edot~$= 4\pi^2I \dot P/P^3 = 5.1 \times 10^{37}\,I_{45}$ ergs s$^{-1}$ 
($I_{45}$ is the moment of inertia in units of 10$^{45}$ g cm$^{2}$) makes \psr~one of most energetic Galactic 
pulsars known, a close tie with several recently reported energetic rotation-powered pulsars 
\citep{c:camilo09,c:gotthelf09}. In the dipole pulsar model, the characteristic age $\tau_c = P/2\dot P = 12.7$ 
kyr, assuming $P \gg P_0$ (the initial PSR period), and the surface dipole magnetic field 
$B_{\rm s} = 1.1 \times 10^{12}$ G. Based on the associated PWN and SNR, its estimated luminosity, and its spin-down 
properties, \psr~is clearly a Crab-like, young energetic rotation-powered PSR responsible for the observed PWN.

\begin{deluxetable}{ll}
\tabletypesize{\footnotesize}

\tablecaption{Measured and derived parameters for \psr~\label{t:ephem}}
\tablewidth{0pc}
\tablehead{ \colhead{Parameter}   & \colhead{Value}   }
\startdata
R.A. (J2000)\dotfill                                    & $14^{\rm h}00^{\rm m}45.69^{\rm s}$ \\
Decl. (J2000)\dotfill                                   & $-63\arcdeg25'42.6^{\prime\prime}$  \\
Period, $P_1$ (ms) @ MJD 54738 \dotfill                 & 31.180373(2)                        \\
Period, $P_2$ (ms) @ MJD 55105 \dotfill                 & 31.1816011(3)                       \\
Period derivative, $\dot P$\dotfill                     & $3.890(6)\times 10^{-14}$           \\
Dispersion measure (cm$^{-3}$~pc)\dotfill               & $563 \pm 4$                         \\
Flux density at 10\un{cm} ($\mu$Jy)\dotfill             & $\sim$ 110                          \\
Flux density at 20\un{cm} ($\mu$Jy)\dotfill             & $\sim$ 250                          \\
Flux density at 50\un{cm} ($\mu$Jy)\dotfill             & $< 500$                             \\
\hline
Characteristic age, $\tau_c$ (kyr)\dotfill              & $12.7$                              \\
Spin-down luminosity, $\dot E$ (ergs\,s$^{-1}$)\dotfill & $5.1\times10^{37}$                  \\
Surface dipole magnetic field, $B_s$ (G)\dotfill        & $1.1 \times10^{12}$                 \\
DM-based distance (kpc)\dotfill                         & $10 \pm 3$                           
\enddata

\tablecomments{\footnotesize $1\sigma$ uncertainties given.}
\end{deluxetable}

\begin{figure}[!htb]
\hspace{0.2cm}
\begin{center}

\includegraphics[height=0.88\linewidth,angle=270,clip=true]{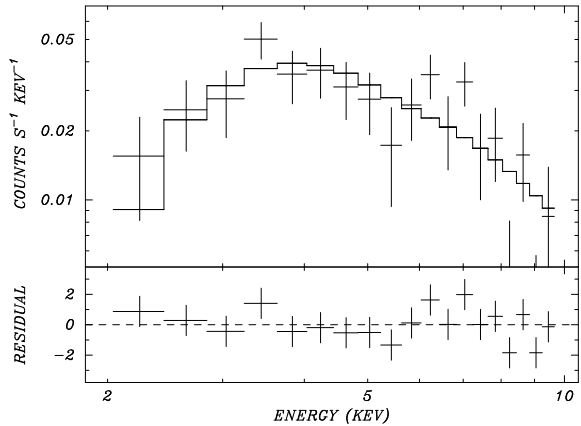}

\caption{\psr~pulsed spectrum in the \xte~PCA 2--20\un{keV} band fitted to
         an absorbed power-law model given in the text. The residuals from the
         best fit model are show. The off-peak spectrum is used as background 
         (see text). \label{f:rxte2}}

\end{center}
\end{figure}

The pulsed emission of \psr~is well isolated in phase with nearly constant emission in-between. This allowed us 
to take advantage of phase-resolved spectroscopy to obtain a spectrum of the pulsed flux using the off-pulse 
emission as a near perfect representation of the background plus unpulsed spectrum. We generated phase-resolved
spectra in 20 phase bins using the \xte~FTOOL {\it fasebin}, aligning the two observations to a common phase 
and summing the result. From this combined spectral file, we generated on- and off-pulsed spectra as input into 
the {\tt XSPEC} spectral fitting package. The response matrix for the phase-resolved spectra was generated using 
the same photon file, selecting only photons from the top layer of each PCU (LR1) to match the {\it fasebin} 
output. We divided the data in to two phases, 0.5~cycles apart to characterise the on-pulse signal and the 
off-pulse signal, respectively. We fitted the on-pulse spectrum with an absorbed power-law spectral model over 
the 2--10\un{keV} band, outside of which the background dominated the pulsed signal. The column density was 
fixed to the \chandra~value as \xte, with its higher energy band, is unable to constain this parameter. For an 
N$_{\rm H}$ = 2.09 $\times 10^{22}$~cm$^{-2}$ the best-fit photon index is $2.0^{+0.5}_{-0.3}$ with a 
$\chi^2 = 1.2$ for $17$ degrees-of-freedom (see Fig.~\ref{f:rxte2}). The measured 2--10\un{keV} pulsed flux 
is $3.0 \times 10^{-13}$~ergs~cm$^{-2}$~s$^{-1}$ (absorbed), and represents $\sim$ 18\% of the total flux 
from the \chandra~source.

\subsubsection{Spectral analysis}

In order to extract the total \xte/PCA spectrum, dominated by the PWN, we also accumulated spectra restricted 
to the top layer of PCU~2. The background was estimated using the faint model. Given the sky position of \igr, 
\xray~emission from the Galactic ridge can contribute to the PCA spectrum. Following previous works 
\citep[e.g.][]{c:prat08,c:rodriguez09}, we estimated its level according to the \citet{c:valinia98} model, 
and by assuming that \chandra~and \xte/PCA are perfectly cross-calibrated. After removing the Galactic 
\xray~contribution, we found that a power-law fit to the 3--20\un{keV} PCA spectrum gives a \nh~similar to 
that found for the three \chandra~components, though not well constrained, (2.25 $\pm$ 1.1) $\times$ 10$^{22}$ 
cm$^{-2}$. Freezing it to the value measured with \chandra~leads to a spectral index of 2.19 $\pm$ 0.10 and 
an unabsorbed 3--20\un{keV} flux of (1.57 $\pm$ 0.3) $\times$ 10$^{-11}$ ergs cm$^{-2}$ s$^{-1}$. The
spectrum measured with PCA turns out to be softer than that measured with \chandra/ACIS in the 2--10\un{keV} band. 
This can be explained by the presence of a break or steepening at high energies in the PWN spectrum. Thus, we 
also fitted the PCA spectrum with a broken power-law, with \nh~and a photon index before the break frozen to the 
\chandra~values, and found a marginal evidence for such a break at $\sim$ 3\un{keV}, with a photon index at higher 
energies of 2.21 $\pm$ 0.09, though with a slightly larger $\chi^{2}$. This emission detected by \xte/PCA is most 
likely related to \snr, and mainly to the PWN diffuse emission, since no other significant \xray~source lies 
within the PCA~field-of-view, at 2\d~from \igr.

\subsection{\integ~\ibisg}
\label{s:ibis}

We analyzed all the public \ibisg~data, from revolution 46 to 713, at less than 12\d~from \igr, with the 
Off-Line Scientific Analysis (OSA) software \citep{c:goldwurm03}, version 7.0. After removing all the noisy 
pointings, as defined in \citet{c:renaud06}, the total effective time amounts to $\sim$ 3.5\un{Ms}. Mosaic 
images were produced in six energy bands from 18 to 65\un{keV} with an additional band between 82 and 
150\un{keV}, in order to extract the non-thermal continuum source spectrum. Furthermore, the very small 
apparent size of \snr~($\Phi$ $\sim$ 2.3\m~= 4.7 d$_{\rm 7}$ pc, with d = 7 d$_{\rm 7}$ kpc) is reminiscent 
of a young SNR. To search for evidence of \ti{Ti} line emission at 67.9 and 78.4\un{keV}, expected from a 
young remnant, we also generated mosaic images in the 65--71, 71--75 and 75--82\un{keV} energy bands. 
These lines emerge from the radioactive decay of the short-lived \ti{Ti} nucleus (with a weighted-average 
lifetime of 85 $\pm$ 0.4 yr, see Ahmad \etal 2006, and references therein). Although firmly detected in 
only the $\sim$ 330 yr old Cas~A SNR \citep{c:iyudin94,c:vink01,c:renaud06}, this radioactive element 
can provide invaluable constraints on the key parameters of the SN explosion.

In the 18--50\un{keV} mosaic image, the best-fit position of \igr~is found to be consistent with the results 
obtained by \cite{c:bird07}, and compatible with a point-like source concident with \snr. The continuum 
emission is safely detected with \ibisg~up to 65\un{keV} and the spectrum is fit by a power-law with a 
photon index of 2.33 $\pm$ 0.29 and a 20--100\un{keV} flux of (1.24 $\pm$ 0.17) $\times$ 10$^{-11}$ ergs 
cm$^{-2}$ s$^{-1}$. No evidence of \ti{Ti} line emission was detected in the direction of \igr, and we set a 
3 $\sigma$ upper limit of 1.5 $\times$ 10$^{-5}$ ph cm$^{-2}$ s$^{-1}$, after combining the two \ti{Ti} 
energy bands.

\begin{deluxetable*}{lcccccccccc}
\tabletypesize{\footnotesize}
\tablecaption{Imaging-spectroscopy of \snr, \psr~and its pulsar wind nebula\label{t:fitres}}
\tablewidth{0pc}
\tablehead{
\colhead{} & \multicolumn{4}{c}{\tt power-law} & \multicolumn{6}{c}{\tt broken power-law} \\
\colhead{} & \colhead{$N_{\rm H}$\tablenotemark{a}} & \colhead{$\Gamma$} & \colhead{Flux\tablenotemark{b}} & \colhead{$\chi^{2}$/$\nu$} & \colhead{$N_{\rm H}$\tablenotemark{a}} & \colhead{$\Gamma_{\rm 1}$} & \colhead{E$_{\rm b}$\tablenotemark{c}} & \colhead{$\Gamma_{\rm 2}$} & \colhead{Flux\tablenotemark{b}} & \colhead{$\chi^{2}$/$\nu$}}
\startdata
PSR (pulsed) RXTE      & 2.09 (--)   &  2.0 (0.40) & 0.36 (0.2) &  20.4/17   & -- & -- & -- & -- & -- & -- \\
PSR (total) Chandra    & 2.09 (0.12) & 1.22 (0.15) & 1.95 (0.5) & 182.5/191  & -- & -- & -- & -- & -- & -- \\
SNR Chandra            & 2.09 (0.12) & 2.56 (0.18) & 1.00 (0.2) & 182.5/191  & -- & -- & -- & -- & -- & -- \\
PWN Chandra            & 2.09 (0.12) & 1.83 (0.09) & 15.1 (2.0) & 182.5/191  & -- & -- & -- & -- & -- & -- \\
RXTE/PCA               & 2.09 (--)   & 2.19 (0.10) & 15.7 (3.0) &  31.8/30   & 2.09 (--) & 1.83 (--) & 3.0 (2.0) & 2.21 (0.09) & 15.7 (3.5) & 31.8/29 \\
IBIS/ISGRI             &      --     & 2.33 (0.29) & 12.4 (1.7)  &   7.2/7   & -- & -- & -- & -- & -- & -- \\
Total\tablenotemark{d} & 2.46 (0.08) & 2.16 (0.03) & 13.4 (0.8)  & 272.6/246 & 2.13 (0.13) & 1.90 (0.11) & 5.6 (0.7) & 2.30 (0.07) & 10.6 (1.5) & 257.1/244 \\
PWN\tablenotemark{e}   &      --     &      --     &      --     &    --     & 2.13 (0.14) & 1.90 (0.10) & 6.0 (0.5) & 2.59 (0.11) &  5.3 (0.8) & 257.1/244 \\
\enddata
\vspace{-0.2cm}
\tablecomments{The uncertainties quoted are at the 68 \% confidence level.}
\tablenotetext{a}{ In units of 10$^{22}$ cm$^{-2}$, assuming the Anders \& Grevesse (1989) abundances}
\tablenotetext{b}{ 2--10 (\chandra), 3--20 (\xte/PCA), and 20--100\un{keV} (\ibisg, Total 
and PWN) unabsorbed fluxes, in units of 10$^{-12}$ $\mathrm{erg\,cm^{-2}\,s^{-1}}$}
\tablenotetext{c}{ Break energy in units of keV}
\tablenotetext{d}{ Total spectrum in the 0.8--100\un{keV} energy range}
\tablenotetext{e}{ PWN spectrum in the 0.8--100\un{keV} energy range, after removing the PSR and SNR contributions} 

\end{deluxetable*}

\subsection{Spectral properties of \snr}
\label{s:spectrum}

\begin{figure}[!htb]
\begin{center}

\includegraphics[width=0.95\linewidth,angle=0]{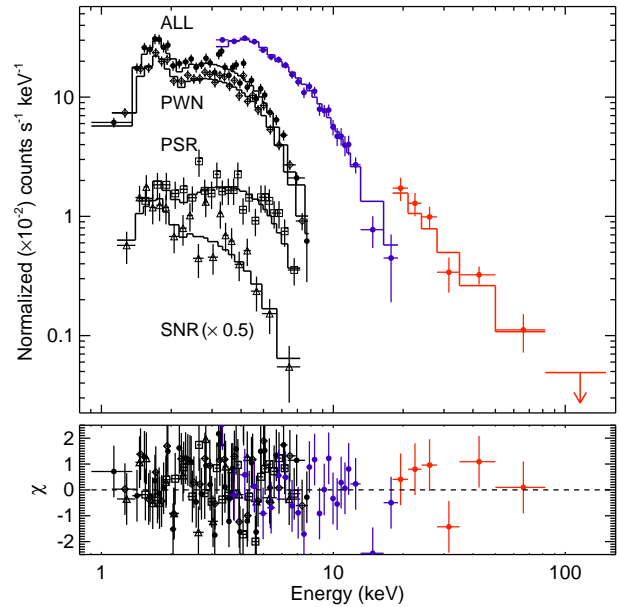}

\caption{Spectra of the different components of \snr~measured with \chandra~(in black). The SNR spectrum 
was divided by 2 for sake of clarity. The \xte/PCA and \integ/\ibisg~spectra are shown in blue and red, 
respectively. Solid lines represent the best-fit models (see Table \ref{t:fitres}). An absorbed power-law 
was used to fit simultaneously the individual \chandra~components, while an absorbed broken power-law 
was fit to the \chandra~spectrum of the whole \xray~emission, together with the \xte/PCA and \ibisg~spectra. 
Residuals are shown in the lower panel. \label{f:spectrum}}

\end{center}
\end{figure}

Spectra of the different components measured with \chandra, together with those of \xte/PCA and \ibisg~are 
shown in Figure \ref{f:spectrum}, and spectral model fits are reported in Table \ref{t:fitres}. In the 
2--10\un{keV} band, the PWN accounts to $\sim$ 85 \% of the whole emission. A simultaneous fit on the 
three spectra (Chandra total emission, \xte/PCA and \ibisg) indicates that a broken power-law is better 
at describing the 0.8--100\un{keV} total emission. According to the F-test, this model is favored over a pure 
power-law at the 3.5 $\sigma$ confidence level (probability of 2 $\times$ 10$^{-4}$). The break energy is 
measured at 5.6 $\pm$ 0.7\un{keV}, and the photon index at higher energies nicely matches that measured 
by \ibisg. We also performed a spectro-imaging analysis of the PWN, in order to search for any spectral 
softening at larger distances from the PSR, as expected from efficient synchrotron burn-off of high-energy 
electrons accelerated in the central regions \citep[see \eg][in the case of G21.5$-$0.9]{c:slane00}. We 
fixed the \nh~to the best-fit \chandra~value, and found a 2.5 $\sigma$ evidence of a spectral softening, 
from $\Gamma$ = 1.68 $\pm$ 0.09 at 5\s~from the PSR, to 2.09 $\pm$ 0.09 at the edge of the nebula. In 
order to measure the 0.8--100\un{keV} PWN spectrum, we first constrained the PSR spectral index and 
2--10\un{keV} flux to be $\sim$ 1.4 and 1.7 $\times$ 10$^{-12}$ ergs cm$^{-2}$ s$^{-1}$ {\it at most}, 
in order not to violate the \ibisg~3 $\sigma$ upper limit in the 82--150\un{keV} energy range, while 
being still consistent with the \chandra~values, at the 1 $\sigma$ confidence level. Then, we 
performed a second simultaneous fit on the three spectra (Chandra total emission, \xte/PCA and \ibisg), 
by freezing the 0.8--10\un{keV} PSR and SNR spectra to the best-fit values, and extracted the {\it minimum} 
0.8--100\un{keV} spectrum of the PWN, whose parameters are presented in Table \ref{t:fitres} (last row).
We find that a broken power-law fit gives a break energy of 6.0 $\pm$ 0.5\un{keV}, in agreement with
that measured above. Due to the hard PSR spectrum, assumed to extend up $\sim$ 100\un{keV} with the same 
spectral index, the {\it minimum} PWN spectrum at higher energies is softer than the total emission, but 
statistically consistent with the spectrum measured by \ibisg.

\subsection{Radio}
\label{s:radio}

\subsubsection{pulsar}

\begin{figure}[!htb]
\begin{center}

\includegraphics[height=0.95\linewidth,angle=270]{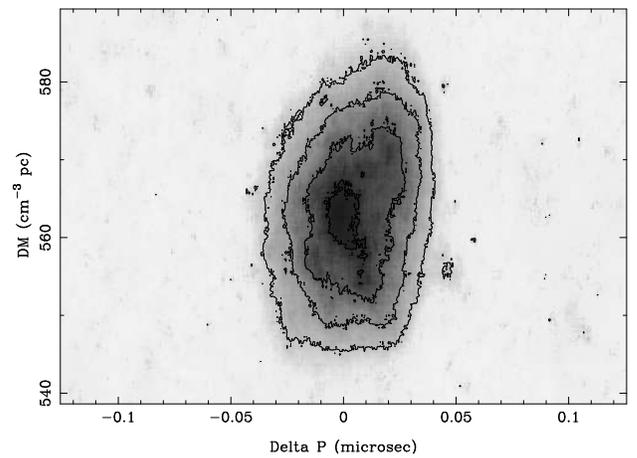}

\caption{Plot of the 20\un{cm} radio pulse S/N ratio in the pulse period -- DM plane around the detection values 
from the 2009 November Parkes observations. The greyscale is linear between S/N values of 0 and 10. S/N contours at 
5, 6, 7 and 8 are overlaid. \label{f:radiopsr1}}

\end{center}
\end{figure}

A search for a radio counterpart to the \xray~pulsar was carried out using the Parkes 64-m radio telescope of the 
Australia Telescope National Facility, CSIRO Astronomy and Space Science, in the 10, 20 and 50\un{cm} bands. 
Observations were made with the centre beam of the 20\un{cm} Multibeam receiver and analogue filterbank system 
\citep{c:mlc01} with sampling interval of 250 $\mu$s on 2009 October 11 (2~h) and 2009 November 26 (4~h), and the 
10--50\un{cm} receiver and Parkes digital filterbank systems (1~h at each of 10\un{cm} and 50\un{cm}), on 2009 
October 13. The received centre frequencies were 3100, 1374 and 732\un{MHz} at 10, 20 and 50\un{cm} 
respectively, and the corresponding bandwidths were 1024, 256 and 64\un{MHz}. A search of the October data
over a period range of $\pm 60\;\mu$s about the nominal pulsar period (see Table~\ref{t:ephem}) and a 
dispersion measure (DM) range of 0 to 1500 cm$^{-3}$~pc yielded some possible candidates with DMs in the 
range 400 -- 600 cm$^{-3}$~pc and signal-to-noise (S/N) ratios of $\lesssim$ 6.5. A similar analysis of 
the longer 20\un{cm} observation in November gave a positive detection with S/N ratio of 8.8 at the 
predicted pulse period and at a DM of $563 \pm 4$ cm$^{-3}$~pc, as shown in Figure~\ref{f:radiopsr1}. 
This detection corresponds to one of the candidates from the analysis of the October 20\un{cm} observations. 

\begin{figure}[!htb]
\begin{center}

\includegraphics[height=0.95\linewidth,angle=270]{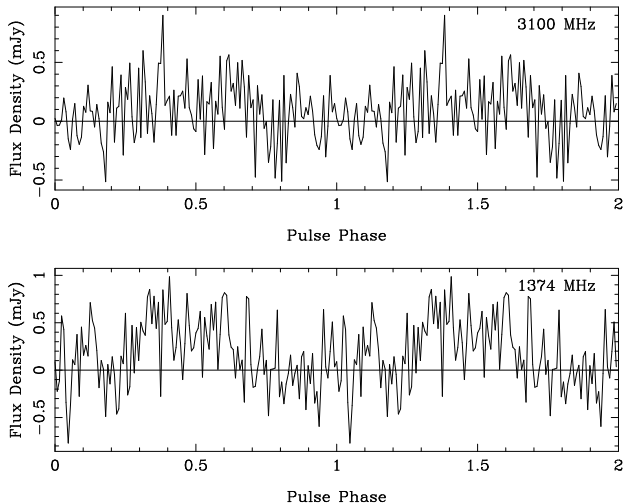}

\caption{Radio pulse profiles at 10\un{cm} (upper) and 20\un{cm} (lower) folded at the predicted pulse period 
and dedispersed with a DM of 563 cm$^{-3}$~pc. The pulse has been approximately centered in both plots. 
\label{f:radiopsr2}}

\end{center}
\end{figure}

Figure~\ref{f:radiopsr2} shows the 20\un{cm} (1374\un{MHz}) pulse profile from the November observations 
and the 10\un{cm} (3100\un{MHz}) pulse profile obtained by folding and dedispersing the data with the nominal 
pulsar parameters and a DM of 563~cm$^{-3}$~pc. The pulsed emission is clearly detected at 10\un{cm} although 
the S/N ratio is low. The profile appears very broad and almost sinusoidal at both frequencies. The profiles 
have been calibrated in flux density units, giving mean pulsed flux densities of approximately 0.25\un{mJy} 
and 0.11\un{mJy} at 20\un{cm} and 10\un{cm}, respectively. The upper limit on the pulsed flux density at
50\un{cm} is about 0.5\un{mJy}. At face value, these results imply a rather flat pulsar spectrum with spectral 
index $\alpha \sim -1.0$ (throughout this paper, S$_{\nu} \propto \nu^{\alpha}$). Radio observations with improved 
sensitivity and with contemporaneous \xray~observations will be required to explore the details of the radio pulse 
morphology and the phase relationship with the \xray~pulse. Distance estimates based on the observed DM are 
9.4\un{kpc} \citep{c:tc93} and 11.3\un{kpc} \citep[NE2001,][]{c:cl02}. Given the large uncertainties inherent in 
DM-based distances, a reasonable estimate for the pulsar distance is $10 \pm 3$~kpc. 

\subsubsection{wind nebula}

Young PSRs are often associated with wind nebulae, usually detected in the radio and \xray~domains. 
We searched for the radio counterpart of the \chandra~SNR in the archival data of the Molonglo Galactic 
Plane Survey at 843\un{MHz} (MGPS-2, Murphy \etal 2007), the 2.4\un{GHz} Parkes survey of the Southern 
Galactic Plane \citep{c:duncan95}, and the Parkes-MIT-NRAO (PMN) survey at 4.85\un{GHz} 
\citep{c:griffith93,c:condon93}. We found an extended MGPS-2 source, MGPS~J140045$-$632542, located at 
R.A.(J2000) = 14\hh00\mm45.17\sec, Decl.(J2000) = $-$63\d25\m42.2\s, 3.5\s~away from the \chandra~point-like 
source. The intrinsic source size of 44.8 $\times$ 31.2\s~(after deconvolution from the MOST beam size 
of $\sim$ 45\s) is depicted in Figure \ref{f:ima_chandra} (left) by the dashed ellipse, which nicely 
matches the \xray~PWN, and thus corresponds to its radio counterpart. The 843\un{MHz} flux density was 
measured to be 217.4 $\pm$ 9.4\un{mJy} \citep{c:murphy07}. In the 4.85\un{GHz} PMN data, we found an 
other cataloged source coincident with the \chandra~SNR. According to \cite{c:wright94}, PMN~J1400$-$6325 
lies at R.A.(J2000) = 14\hh00\mm45.2\sec, Decl.(J2000) = $-$63\d25\m43\s. With an angular resolution of 
5\m~(FWHM), the source is unresolved, and features a flux density of 113 (131) $\pm$ 10\un{mJy}, 
depending on the fit procedure used \citep[see][]{c:griffith93,c:wright94}. No significant emission is 
detected in the Parkes image at 2.4\un{GHz} \citep{c:duncan95}, and we calculated a 3 $\sigma$ upper 
limit of 0.6\un{Jy}. A power-law fit on these three radio measurements gives a spectral index $\alpha$ of 
$-$0.33 $\pm$ 0.05, $-$0.205 $\pm$ 0.008, and a flux density at 1\un{GHz} of (0.207--0.204) $\pm$ 0.008\un{Jy}, 
depending on the adopted MGPS-2 source flux. Such $\alpha$ index is similar to those measured in other radio 
PWNe \citep{c:gs06}, and we can safely conclude these MGPS-2 and PMN sources are the counterparts of the 
\xray~wind nebula discovered with \chandra.

No radio shell-type structure is detected in the MGPS-2 image, even though the MOST angular resolution 
should allow us to separate it from the PWN. We have calculated an upper limit on the SNR surface 
brightness by modelling the SNR as a uniform annulus with a mean radius of 1.15\m~and a width of 
0.15\m~(see Fig. \ref{f:ima_chandra}, right) and convolving it with the MOST beam size. The upper 
limit, derived by summing all the pixels within a region encompassing the FWHM of the simulated MGPS-2 
shell emission, is 2.5 $\times$ 10$^{-21}$ W m$^{-2}$ Hz$^{-1}$ sr $^{-1}$ at 843\un{MHz}, which 
corresponds to 2.3 $\times$ 10$^{-21}$ W m$^{-2}$ Hz$^{-1}$ sr $^{-1}$ at 1\un{GHz} (for $\alpha$ = $-$0.5), 
as commonly used in studies of radio shell-type SNRs \citep{c:green09}.

\section{Discussion}
\label{s:discu}

\subsection{General considerations}
\label{s:gnrl}

After the Crab pulsar, \psr~is one of the most energetic pulsars known in the Galaxy, rivaling the recently 
discovered PSRs associated with G12.82$-$0.02 and G0.9+0.1 SNRs \citep{c:gotthelf09,c:camilo09}. Its characteristic 
age $\tau_c$ of 12.7\un{kyr} likely overestimates the true age $\tau$, based on comparison with the SNR estimate 
(see below), consistent with the fact that the pulsar has not spun down significantly. We then assume in the 
following that $\tau \ll \tau_c$ (\ie~P $\sim$ P$_{0}$ and \edot~$\sim$ \edot$_{0}$).
The commonly used relations between PSR/PWN \xray~spectra (2--10\un{keV} indices and luminosities) and the 
intrinsic PSR properties \citep{c:possenti02,c:gotthelf03,c:mattana08,c:li08}, provide 
consistent estimates with those measured here, though with the large uncertainties inherent of such correlation 
studies. Amongst these relations, those between PSR/PWN \xray~luminosities and \edot~point towards a distance to the 
source of 6-10\un{kpc}, compatible with the DM-based estimate of 10 $\pm$ 3\un{kpc}. On one hand, the high 
\nh~derived from the \chandra~spectra suggests either a large source distance, or local absorption that 
could be undetectable in the existing large-scale HI and $^{12}$CO Galactic surveys at moderate angular resolutions. 
Using the latest relation between optical extinction and \nh~of \citet{c:guver09}, we find an extinction A$_{\rm V} \sim $ 
9 toward \igr. Furthermore, according to the Galactic interstellar extinction map of \citet{c:marshall06}, 
such A$_{\rm V}$ along the \igr~line of sight corresponds to a distance of 5-8\un{kpc}. For such a high extinction, 
only in the case of an over-luminous core-collapse SN event (with M$_{\rm V}$ $< -$17, see Richardson 2009), would 
the stellar explosion of \snr~have been visible with naked eye (m$_{\rm V}$ $\leq$ 6) for distances $\geq$ 5\un{kpc}. 
On the other hand, its Galactic latitude $b \sim$ $-$1.6\d~translates into an height from the Galactic Plane of $\sim$ 
280\un{pc} at 10\un{kpc}, which is much larger than the typical scale height of the Galactic molecular disk 
\citep[$\sim$ 55\un{pc}, \eg][]{c:ferriere01} within which core-collapse SNe are known to usually explode\footnote{We note
that \snr~could potentially originate from a runaway OB / Wolf-Rayet star \citep[see \eg][]{c:dray05}. This scenario
would explain the large height from the Galactic Plane for a distance of 10\un{kpc}.}. We also note that the Galactic 
longitude of \snr~corresponds to the Crux-Scutum arm tangent, at 6-7\un{kpc} \citep{c:vallee08}, where much of the molecular 
material along this line of sight lies. In order to account for all the above-mentioned estimates with their respective
uncertainties, we then adopt a distance of 7\un{kpc}, which implies a height similar to that of the Crab nebula, and scale 
the distance in terms of $d_7$ = d/7\un{kpc}.

The very small angular size of \snr~$\sim$ 2.3\m~= 4.7$d_{7}$ pc already places it amongst the smallest 
known SNRs, and could in turn imply a very young age, similar to the recently rediscovered G1.9+0.3 SNR 
\citep{c:reynolds08}, and the well known Tycho, Kepler and Cas~A SNRs. For a uniform ambient medium of density 
$n_0$\,cm$^{-3}$, the SNR has swept up only M$_{\rm sw} = (4 \pi / 3)\,n_0 R^3 \sim 1.3\,n_0 d_{7}^{3}$ \msun~and 
thus may still be in free expansion phase. In such case, for a typical ejected mass of M$_{\rm ej} = 10\,M_{\rm 10}$\,\msun~and 
a SN explosion kinetic energy of $10^{51}\,E_{51}$\,ergs, the expansion velocity would be 
$\sim$ $3200\,(E_{51}/M_{\rm 10})^{1/2}$\,\kms, implying an age $\sim$ 720 $\,d_{7} (M_{\rm 10}/E_{51})^{1/2}$ yr. Such 
a young age is also supported by the nearly circular shape of the SNR, with the PSR lying very near its geometrical center. 
An offset of $\lesssim$ 2.5\s, as estimated on the \chandra~image, implies a projected space PSR velocity of only 
$\lesssim$ 80 $\,d_{7} \tau^{-1}_3$ \kms, with $\tau = 10^3\,\tau_3$ yr, \ie~significantly smaller than the average found by 
\cite{c:hobbs05} for young PSRs. If we assume \snr~has produced the same amount of \ti{Ti} as Cas~A \citep{c:renaud06}, the 
\ibisg~upper limit on the two low-energy \ti{Ti} lines would translate into a lower limit on its age of $\sim$ 280\un{yr} at 
7\un{kpc}. If younger, \snr~would not have originated from one of these \ti{Ti}-rich SN explosions whose remnants (except 
Cas~A) have been unfruitfully searched by several missions over the past decades \citep[\eg][]{c:renaud04,c:the06}.

\subsection{PWN broadband emission and energetics}
\label{s:energetics}

In order to account for the different spectral indexes measured in radio and \xray~domains (see section 
\ref{s:radio} and Table \ref{t:fitres}), the PWN broadband synchrotron spectrum should exhibit a break at 
$\sim$ 3 $\times$ 10$^{4}$ GHz. Such frequency, if originated from synchrotron cooling of accelerated electrons inside 
the PWN, would correspond to an uncomfortably high magnetic field strength $B \sim 330\,\tau_3^{-2/3}$ $\mu$G. In such 
case, the PWN should have appeared smaller in \xrays~than in radio, in contradiction with what is shown in Figure 
\ref{f:ima_chandra}. This break could instead reflect the injected electron spectrum \citep[see \eg][]{c:camilo06} 
or the nature of the magnetic field and of the resulting synchrotron emission \citep{c:fleishman07}. As presented in 
section \ref{s:spectrum}, the break measured at $\sim$ 6\un{keV} could be that expected from synchrotron cooling and, 
thus, provides a PWN age $\tau$ $\sim 880\,B_{10}^{-3/2}$ yr, with $B = 10\,B_{10}$ $\mu$G. VHE observations 
can help further constrain the magnetic field strength inside the PWN. As reported by \cite{c:chaves08}, no 
point-like source toward \igr~is detected above 5 $\sigma$ in the latest map of the \hess~Galactic Plane Survey, 
in 15\un{hours} of effective time. This translates into a \hess~upper limit of 4 \% of the Crab nebula 
\citep{c:khelifi08}, \ie~3 $\times$ 10$^{-12}$ ergs cm$^{-2}$ s$^{-1}$ in the 1--10\un{TeV} energy range. 
In a simple one-zone leptonic scenario, the VHE emission comes from inverse-Compton scattering of VHE electrons 
on ambient photons (CMB and infrared and optical Galactic fields, from dust and stars, respectively). From the 
ratio between the \xray~flux and the \hess~upper limit, we find that the magnetic field must be larger than 6 
$\mu$G, and then, the age must be smaller than $\sim$ 1900\un{yr}. The corresponding maximal energy content in 
particles E$_{\rm p, max}$ amounts to $\sim 2.2\,\times\,10^{48}\,d_{7}^{2}$ ergs.

PWN energetics can also provide useful insight on the source age, as young PWNe usually feature a ratio 
between the internal energy $E_{\rm int}$ and the PSR rotational spin-down energy $E_{sd} \equiv$ \edot$\tau$ $<$ 1 
\citep[$\sim$ 0.44 for $\tau \ll$ the initial PSR spin-down time $\tau_0$, and depends weakly on the supernova 
density profile, see][equation 30]{c:chevalier05}. On one hand, E$_{\rm p, max}$ thus provides a conservative limit 
E$_{sd, max}$ $\sim 5\,\times\,10^{48}\,d_{7}^{2}$ ergs, which implies a maximal age of $\sim$ 3000\un{yr} for 
\edot~$= 5.1 \times 10^{37}$ ergs s$^{-1}$. On the other hand, a lower limit on $E_{\rm int}$ can be obtained 
from equipartition arguments \citep{c:govoni04,c:chevalier05}. We estimate that 
$E_{\rm int, min} \sim 4\,\times 10^{46}\,d_7^{17/7}$ ergs, \ie~$\ll$ \edot$\tau_3 = 1.6\,\times\,10^{48}$ ergs, 
and a magnetic field at equipartition B$_{\rm eq} \sim 81\,d_7^{-4/14}$ $\mu$G, from the radio properties of the PWN. 
In case \igr~is at the equipartition between particles and magnetic field, its age would then be $\ll$ 1\un{kyr}. However, 
\hess~has recently discovered VHE $\gamma$-ray emission toward two other young PWNe, G21.5$-$0.9 and Kes~75 
\citep{c:djannati08}. These observations imply magnetic fields of 10--15 $\mu$G, well below the equipartition values, 
and therefore support earlier suggestions \citep{c:chevalier05} that some young PWNe are particle dominated. If the 
same applies to \igr, $E_{\rm int}$ at the equipartition does not reflect the whole PWN energy content, and 
$E_{\rm int}({\rm B}) \propto {\rm B}^{-3/2}$ in the particle-dominated regime. In this regime, following 
\cite{c:chevalier05}, the $E_{\rm int}(d,{\rm B})$/\edot$\tau \sim$ 0.44 relation, valid when $\tau \ll \tau_{0}$ as 
assumed here, would be consistent with the existence of a break measured at $\sim$ 6\un{keV} ($\tau$ 
$\sim 880\,B_{10}^{-3/2}$ yr) for a distance of 7--8\un{kpc}.

\subsection{The evolutionary stage of \snr}
\label{s:evolsnr}

As discussed previously, several arguments point towards a very young composite SNR. The ratio\footnote{The radio PWN 
found in the MGPS-2 data is elliptical, so that we suppose its effective size to be $\sqrt{44.8 \times 31.2}$\s. This 
yields a PWN to SNR size ratio of $\sim$ 37\s/2.3\m~$\sim$ 1/4.} between the PWN and SNR radii is \textsl{R} 
$\approx$ 1/4, and can be used to probe the parameters governing the early evolution of composite SNRs, namely 
$E_{51}$, $M_{\rm ej}$ and $n_0$ \citep{c:vds01,c:blondin01}, given the \edot~measured with \xte. First, $n_0$ 
and $E_{51}$ are constrained by the faint \xray~level, and the 1\un{GHz} upper limit, of the SNR synchrotron surface 
brightness. According to \cite{c:bv04}, such low values can only be explained by a very tenuous medium 
($n_0$ $\lesssim$ 0.01 cm$^{-3}$)\footnote{This argues against a local absorption at the origin of the large
\nh, and distances much less than the Crux-Scutum spiral arm tangent.}, together with a sub-energetic explosion. 
Furthermore, using the so-called radius method of \cite{c:vds_wu01}, the rotational energy $E_{sd}$ relates to the PWN 
to SNR radius ratio \textsl{R} as: $E_{sd} \sim 2 \times 10^{48}\,E_{51}\,(\textsl{R}/0.25)^3\,(\eta_3/2)^{-3}$ 
erg, or $\tau \sim 1200\,E_{51}\,(\textsl{R}/0.25)^3\,(\eta_3/2)^{-3}$ yr, with $\eta_3$ a dimensionless parameter 
ranging between 1 and 3 during the early evolution phase (see their Fig. 1). With the upper limit on $E_{sd}$ 
estimated from the \hess~non detection, and the lower limit $E_{\rm int, min}$ from equipartition arguments, we 
infer $0.02\,d_7^{17/7}\,(\eta_3/2)^3 <$ $E_{51}$ $< 1\,d_7^{2}\,(\eta_3/2)^3$, in agreement with the weak radio 
and \xray~surface brightnesses of the SNR. We estimate the initial PSR period $P_{0} \gtrsim 29.5$\un{ms} (for a 
braking index n = 3), for a current period of 31.18\un{ms}. Based on the \cite{c:blondin01} and \cite{c:chevalier05}
calculations, we then explored a wide range of $E_{51}$, $n_0$ and $M_{\rm ej}$ (the power-law index of the ejecta 
distribution was set to 12). For a distance of 7\un{kpc}, if $n_{0}$ is set to 0.01 cm$^{-3}$, we find that only a 
sub-energetic ($E_{51}$ = 0.05), low ejecta mass (M$_{\rm ej}$ = 3\msun) SN explosion could explain the measured 
\textsl{R}. The resulting SNR age and PWN magnetic field are $\sim$ 920\un{yr} and $\sim$ 10 $\mu$G, respectively. 
Note that the constraint on $n_{0}$ is only {\it qualitative}, as a quantitative estimate would require a detailed 
modelling of the SNR synchrotron emission, as performed by \cite{c:bv04}. If we relax this constraint and set $n_{0}$ to the 
Galactic value of 1 cm$^{-3}$, this would result in a larger energy of the explosion ($E_{51}$ = 0.4), and a smaller 
SNR age ($\tau$ $\sim$ 550\un{yr}), for the same mass of the ejecta. In this regard, knowing the nature of the SNR 
\xray~emission is crucial to assess the thermal flux level, and then, the surrounding density.

\section{Conclusion}

Even though some assumptions have been made in the previous estimates (\eg~constant PWN magnetic field and PSR 
spin-down power), they give valuable insight on the nature of this new Galactic composite SNR. \snr~harbors a highly 
energetic 31.18\un{ms} pulsar, \psr, which powers a wind nebula, both lying at the center of the host shell. All of 
the existing multi-wavelength observations suggest it is a young SNR ($\lesssim$ 10$^{3}$ yr), and most likely distant 
($>$ 5\un{kpc}). However, many questions still remain to be answered. First, the distance is not very
well constrained, though a large distance is favored by the dispersion measure of the radio pulse emission and by the 
large \nh~measured with \chandra. High-resolution observations of the ISM tracers such as HI and $^{12}$CO are then 
warranted to assess the surrounding medium properties. Moreover, the break measured at 6\un{keV} and the spectral softening 
at increasing distances in the PWN need to be confirmed, and the SNR \xray~spectrum needs to be investigated with more 
\xray~data. Nevertheless, \snr~falls into the emerging class of ``multi-wavelength'', young ($\tau \lesssim$ a few 10$^{3}$ 
yr) and composite SNRs, harboring very energetic PSRs and wind nebulae shining in radio, \xrays~and potentially in HE/VHE 
\gammarays. The list includes Kes~75 and G21.5$-$0.9 \citep{c:gotthelf00,c:camilo06,c:bietenholz08,c:djannati08}, and more 
recently G0.9+0.1 \citep{c:aharonian05a} and HESS~J1813$-$178 \citep{c:aharonian05b}, whose long-expected PSRs have recently 
been discovered \citep{c:camilo09,c:gotthelf09}. It is of interest to note that \fermi/LAT has not detected a bright 
source \citep{c:fermi_soucat}, nor pulsar \citep{c:fermi_psr1,c:fermi_psr2} coincident with \snr, although, at first 
glance, energetic PSRs should be the most easily detectable sources. However, not all of the above-mentioned young and 
energetic PSRs have been detected by \fermi. From the first \fermi~catalog of gamma-ray PSRs, the sensitivity for 
a blind search of pulsed emission in the Galactic Plane is conservatively taken to be 2 $\times$ 10$^{-7}$ cm$^{-2}$
s$^{-1}$ above 100\un{MeV} \citep{c:fermi_psr2}. Assuming a spectrum similar to that of PSR~J1833$-$1034 associated 
with G21.5$-$0.9, \psr~features a maximal efficiency $\eta =\,L_{\gamma}$/\edot~of $\sim$ 1 $d_{7}^2$ \%. This is
close to what is measured from other young PSRs (see Fig. 6 of \citet{c:fermi_psr2}), and argues in favor of a fairly
large distance to the source, as outlined in section \ref{s:discu}. Further HE/VHE observations of \snr~will certainly 
provide important constraints both on the PSR gamma-ray spectrum and on the PWN magnetic field strength.

\acknowledgments

We would like to thank Dr. Jean Swank and the \xte~team for making available a ToO, and O. de Jagger, Y. Gallant,
M. Bietenholz and F. Bocchino for helpful discussions. M.R. and F.M. acknowledge the French Space Agency (CNES) for financial 
support. The present work is based on observations with \integ, an ESA project with instruments and science data center 
(ISDC) funded by ESA members states (especially the PI countries: Denmark, France, Germany, Italy, Switzerland, Spain, 
Czech Republic and Poland, and with the participation of Russia and the USA). ISGRI has been realized and maintained in 
flight by CEA-Saclay/DAPNIA with the support of CNES. The Parkes telescope is part of the Australia Telescope which is
funded by the Commonwealth Government for operation as a National Facility managed by CSIRO.

\end{document}